\documentclass{article}
\usepackage{verbatim}
\usepackage{graphicx}
\usepackage{url} 
\usepackage[square, numbers, sort&compress]{natbib}
\usepackage{amsmath}
\usepackage{amssymb}
\usepackage{listings} 
\usepackage{longtable} 

\usepackage{amsthm} 
\usepackage[table,xcdraw]{xcolor} 
\usepackage{tikz} 
\usepackage{chngcntr} 

\title{Proxy Forecasting to Avoid Stochastic Decision Rules in Decision Markets}
\author{Wenlong Wang and Thomas Pfeiffer}
\date{March 2023}

\begin{document}

\maketitle

\begin{abstract}
    Information that is of relevance for decision-making is often distributed, and held by self-interested agents. Decision markets are well-suited mechanisms to elicit such information and aggregate it into conditional forecasts that can be used for decision-making. However, for incentive-compatible elicitation, decision markets rely on stochastic decision rules which entails that sometimes actions have to be taken that have been predicted to be sub-optimal. In this work, we propose three closely related mechanisms that elicit and aggregate information similar to a decision market, but are incentive compatible despite using a deterministic decision rule. Following ideas from peer prediction mechanisms, proxies rather than observed future outcomes are used to score predictions. The first mechanism requires the principal to have her own signal, which is then used as a proxy to elicit information from a group of self-interested agents. The principal then deterministically maps the aggregated forecasts and the proxy to the best possible decision. The second and third mechanisms expand the first to cover a scenario where the principal does not have access to her own signal. The principal offers a partial profit to align the interest of one agent and retrieve its signal as a proxy; or alternatively uses a proper peer prediction mechanism to elicit signals from two agents. Aggregation and decision-making then follow the first mechanism. We evaluate our first mechanism using a multi-agent bandit learning system. The result suggests that the mechanism can train agents to achieve a performance similar to a Bayesian inference model with access to all information held by the agents.
\end{abstract}

\section{Introduction}
Consider a company that has developed two alternative product lines and needs to decide which one to move into production. The decision maker (principal) has a noisy signal about the likelihood of success of the product lines and also engages a group of experts who have their own independent noisy signals. The aim of the decision maker is to make the best possible decision, given the experts’ and her own signals. The experts are self-interested and require incentives to reveal their information. The principal needs to design the incentives such that the experts cannot benefit from manipulative strategies that result in larger rewards for inaccurate information.

Collective decision-making processes need to address two challenges. Firstly, a process needs to incentivise self-interested participants to provide their information accurately. Secondly, multiple pieces of information need to be aggregated and mapped to the final decision. Several options for eliciting advice for decision-making have been discussed in \cite{Chen2014ElicitingMaking}. One option is to elicit forecasts about the consequences of the available actions and make a decision based on these forecasts. An alternative is to directly elicit recommendations about which action to take.

Prediction markets are suitable mechanisms to elicit forecasts from groups of experts when realised outcomes do not depend on the selected actions. However, when the future outcomes depend on the choice of the principal, it is difficult to design proper incentives because the unselected actions become counterfactual, and forecasts about their consequences cannot be evaluated by a strictly proper scoring rule. A simple mechanism where the principal first elicits forecasts about the available actions and then selects the action that is forecasted to be most beneficial is therefore prone to strategic manipulation by self-interested experts \cite{Othman2010DecisionMarkets, Chen2011InformationMaking}. 

As a solution to this problem, proper decision markets have been proposed \cite{Chen2011DecisionIncentives}. Proper decision markets use strictly proper scoring rules to elicit forecasts, and then use a stochastic decision rule to map the elicited forecasts to a probability distribution over the available actions. This probability distribution needs to have full support, i.e., assign each action a non-zero probability of being executed. Forecasts about the available actions can then be rewarded such that the expected payoff for a forecast does not depend on which action is selected.

The stochastic decision rule required to properly incentivise decision markets means that the principal needs to select actions that are forecasted to be suboptimal. This introduces inefficiency in the decision-making process. While the probabilities assigned to suboptimal action can be made arbitrarily small, small probabilities in the decision rule lead to reward distributions with high variance and to large worst-case losses for the principal \cite{Chen2011DecisionIncentives,Wang2022SecuritiesMarkets}.

Naturally, the interest of the principal is to elicit collective forecasts for the available actions that are as accurate as possible, and then deterministically select the action that is forecasted to be the most desirable one. In this paper, we propose a set of mechanisms which fulfil these requirements and thus fill a gap in the literature on collective decision-making. By eliciting forecasts for observable proxies for the consequences of the available actions, these mechanisms separate the scores for experts from the decision made by the principal and, therefore, can use deterministic decision rules. Such a proxy needs to be verifiable and statistically correlated with the desirability of the actions; in the simplest case, this could be a signal held by the principal. The principal then elicits forecasts about this proxy and uses the aggregated reports to deterministically select an action. Our work builds on ideas from peer predictions, where peers provide proxies for an unobservable ground truth. 

We illustrate our mechanisms with simulations of learning in a contextual bandit problem where contextual information (i.e., signals) is distributed over multiple self-interested agents. Such a setting was used previously to study multi-agent learning under a decision market mechanism \cite{Wang2022DecisionProblems}.

This paper is organised as follows: In Section 2, we briefly discuss related work. In Section 3, we propose a mechanism that allows a principal to aggregate distributed information and make the decision deterministically with the assumption that the principal has an independent and identically distributed signal. We then extend the mechanism and describe two variants that work under different assumptions. In Section 4, we show simulation results to illustrate that one mechanism can be used for multi-agent learning. In Section 5, we conclude the work and discuss future directions.

\section{Related Work}
The work presented here is related to work on strictly proper scoring rules, prediction markets, and decision markets. Strictly proper scoring rules for probabilistic forecasts are described in \cite{Gneiting2007StrictlyEstimation}. Prediction markets expand these scoring rules to aggregate information from multiple forecasters \cite{Hanson2007LogarithmicAggregation,Chen2007AMakers,Pennock2007ComputationalMarkets,Chen2010ALearning,Chakraborty2016TradingMarkets,Freeman2017CrowdsourcedMarkets}. Decision markets are mechanisms to aggregate conditional forecasts for decision making, and are described in \cite{Hanson1999DecisionMarkets,Othman2010DecisionMarkets,Chen2011InformationMaking,Chen2011DecisionIncentives}. An application of decision markets to multi-agent contextual bandit system is described in  \citeauthor{Wang2022DecisionProblems}; detailed background on multi-bandit problems is provided in \cite{Lai1985AsymptoticallyRules,Agrawal1995TheProblem,Sutton2018ReinforcementIntroduction,Lai1985AsymptoticallyRules}.
 
The approach to using a proxy to verify forecasts for unobservable outcomes follows a well-established strategy from peer prediction mechanisms \cite{Prelec2004AData,Miller2005ElicitingMethod,Jurca2009MechanismsTruthful,Dasgupta2013CrowdsourcedProficiency,Kong2019AnTruth-telling}. In brief, peer prediction mechanisms incentivise participants to truthfully report a signal they have received. In a real-world application, such a signal could be an experience in a restaurant, which is reported in a review and may serve as a proxy for the restaurant’s quality. Peer prediction mechanisms use the correlation between different agents’ signals to ensure truth-telling is a Nash equilibrium \cite{Miller2005ElicitingMethod,Prelec2004AData}. When agents perform multiple similar tasks, this mechanism can be designed to ensure that agreeing with another reference report is rewarded while blind agreement is penalised \cite{Dasgupta2013CrowdsourcedProficiency}. A number of studies have expanded this mechanism along different dimensions \cite{Jurca2009MechanismsTruthful,Witkowski2012PeerPrior,Witkowski2012APopulations,Zhang2014ElicitabilityPrediction,Liu2017Machine-LearningPrediction,Kong2018WaterInformation,Liu2020SurrogateRules}.

Our work is closely related to \cite{Witkowski2017ProperRules} which proposes proper proxy scoring rules to incentivise a forecaster to truthfully reveal her probabilistic belief with a proxy instead of a future outcome. \citeauthor{Witkowski2017ProperRules} use existing empirical data to evaluate their proxy scoring rules. Our work follows this work to avoid the disadvantages of a stochastic decision rule in proper decision markets. We test our mechanism in simulations of a multi-agent bandit learning system that allows self-interested agents to converge to optimal collective decision-making (more detail in Section \ref{sec:pd_simulation_results}).

\section{Mechanism Design}
This section will outline three closely related mechanisms that use proxies to avoid the disadvantages of stochastic decision-making in proper decision markets. In the first mechanism, we assume that, similar to related work on decision markets \cite{Hanson1999DecisionMarkets,Othman2010DecisionMarkets,Chen2011DecisionIncentives}, a principal decides between multiple alternative actions. Moreover, there is a group of agents each of which receives independent and identically distributed (iid) signals about the likely outcome for one of the actions. Unlike previous work on decision markets, we assume that the principal has her own signal about one of the available actions, which is iid from the other agents’ signals. The principal elicits and aggregates the agents' signals by using her own signal as a proxy, i.e. she elicits forecasts from the agents about her own independent signal, and then selects one of the available actions based on these forecasts. In the second mechanism, the principal does not have an own signal, but recruits one of the agents as an advisor and uses the advisor’s signal to elicit and aggregate the other agents’ signals. Because the principal requires only a single signal to elicit all other signals, she can use a fraction of her own reward as an incentive to align the advisor's interest and retrieve the signal. Thirdly, the principal separates two agents from the rest of the agents, uses peer prediction to elicit the signals from these two agents, and then elicits the remaining agents’ signals following the procedure of the first mechanism. 

For simplicity, we assume for all mechanisms that each action has Bernoulli outcomes (e.g. Success; Failure), one of which is preferred by the principal. As long as probabilistic forecasts for the outcomes allow to generate preference rankings over the actions, the outcome space can be extended to finite, mutually exclusive sets,  as has been considered in existing research \cite{Chen2014ElicitingMaking}. Moreover, signals are assumed to be binary (e.g. 0 or 1), and the probability of receiving a particular signal for an action depends on the Bernoulli outcome for that action. However, in principle more complex signals can be considered as well.

\subsection{Principal with Own Signal}
The principal aims to make an informed decision between alternative actions, and has a signal about the Bernoulli outcome for one of the actions. The principal engages with a group of agents who also have independent signals about an action. To make the best possible decision, the principal needs to incentivise agents to reveal their information and aggregate the elicited information to select an action. Instead of using the observed outcome to reward the agents for accurate forecasts, as is done in decision markets and would require a stochastic decision rule, she uses her own signal as a proxy. Rather than eliciting probabilistic forecasts for the Bernoulli outcomes for each action, agents are incentivised to provide forecasts for her signal. Because the outcome from the action is not required to score the agents, this allows for deterministically selecting an action.  

Similar to prediction and decision markets, agents report a set of probabilities (one probability for the signal being 1 for each action) in sequential order. The probabilistic reports depend on an agent’s signal and the reports from the previous agent in sequence. The first agent uses uninformative odds as prior (i.e., two 50/50s for a two-action scenario), or a common prior. After the last agent makes the reports, the principal will use a strictly proper scoring rule and her signal as a proxy for the corresponding action’s outcome to evaluate all probabilistic reports. As is done in properly incentivised prediction markets and decision markets, the principal assigns a reward to the evaluated report that is equal to the difference between the strictly proper score of the reports and the previous reports to ensure efficiency in assigning rewards \cite{Hanson2003CombinatorialDesign,Chen2011DecisionIncentives}. This mechanism resembles peer prediction, except that the peer evaluation uses the principal’s signal and the principal is assumed to resolve the markets truthfully \cite{Miller2005ElicitingMethod}.  

Evaluating with a proxy guarantees the incentive combability for agents as their rewards do not depend on the principal’s decision. The principal can therefore use a deterministic decision rule that maps her signal and the aggregated reports from the last agent to an action that is predicted to have the highest chance of achieving an outcome desired by the principal. The best deterministic decision rule can be inferred by the principal if all priors and conditional probabilities are known, and can also be learned, as is illustrated in the simulations of the contextual bandit problem, which are detailed in Section 4. 

\subsection{Principal with One Agent Acting as an Advisor}
Previous work has shown that offering a single agent a portion of the principal’s reward provides a simple way to align the agent’s interest with the principal’s interest \cite{Othman2010DecisionMarkets}. Such an alignment allows the principal to deterministically select an action following the agent’s advice \cite{Chen2014ElicitingMaking}. However, how to provide incentives to multiple advisors who might, based on their signals, provide conflicting advice is an open question. 

Because the principal requires only one single signal to elicit all other signals, in cases where the principal does not have access to her own signal, she can incentivise one agent to act as an ‘advisor’. The principal then elicits the advisor's signal, and uses this signal as a proxy for the ground truth. Note that while in \cite{Chen2014ElicitingMaking}, the advisor recommends an action directly according to the received signal, in our mechanism, the truthful announcement of the advisor’s signal is required. Since the advisor’s and principal’s interests are aligned, the advisor can be expected to truthfully provide the signal. The other agents, who are again assumed to have received independent signals, are incentivised to make forecasts about the advisor’s signal. Similar to the procedure described in Section 3.1, the principal then uses these forecasts to deterministically select an action. Note that for this mechanism to function, the advisor cannot be allowed to participate, or collude with the other agents, in the incentivised forecasting. 

\subsection{Principal with Two Agents Evaluated by Peer Prediction}
The mechanism proposed in the previous section lifts the assumption that the principal has an iid signal, similar to the agents. However, to incentivise the advisor, the principal requires the outcome of the selected action to materialise. The mechanism will be more flexible if the iid signals can be acquired without verification, as the outcome of actions may take a long time to materialise. There is substantial research about peer prediction mechanisms eliciting information without verification \cite{Prelec2004AData,Miller2005ElicitingMethod,Jurca2009MechanismsTruthful,Witkowski2012PeerPrior,Dasgupta2013CrowdsourcedProficiency,Zhang2014ElicitabilityPrediction,Liu2017Machine-LearningPrediction,Kong2019AnTruth-telling}.

The setting of the principal and agents follows Section 3.2 in that the agents, but not the principal, have access to iid signals. However, in the third variation of the proposed mechanism, we separate two agents and refer to them as peers. Although any peer mechanism where truthfully announcing signals is a Nash equilibrium satisfies our requirements, the mechanism with the smallest possible number of peers is best suited because peer signals cannot be aggregated in the market. We use the mechanism proposed by \cite{Dasgupta2013CrowdsourcedProficiency} as an example to expand our mechanism with advantages inherited from peer prediction mechanisms. At the procedure's beginning, agents and peers sample the signals. The peers are required to announce the signal to the principal. The principal will evaluate a peer's announcement with the other peer's announcement, and will reward agreement. However, this does not prevent agents from providing uninformative announcements, such as always making identical announcements regardless of the actual signal. We, therefore, follow \cite{Dasgupta2013CrowdsourcedProficiency} and resolve this problem by requiring peers to participate in multiple prior similar tasks to compute a statistic to penalise blind agreements. In our problem, the multi-task requirement can be implemented with a memory of the announcements in previous steps. Again, after the principal retrieves the announcement, which is an iid signal guaranteed by peer prediction mechanisms, the procedure follows Section 3.1 using the signal of one of the peers as a proxy. Note that the principal must ensure that the peers are not colluding with each other, or with the agents who forecast one of their signals.

\section{Simulations for Multi-agent Bandit Learning }
In this section, we describe simulations of multi-agent learning based on the mechanism described in Section 3.1. We model the decision-making mechanism as a multi-agent contextual bandit problem with Bernoulli outcomes. A principal aims to decide among several alternative actions, while signals about the quality of the actions are distributed among a group of self-interested agents. Agents and the principal receive independent and identical distributed (iid) signals for one of the actions, this is in contrast to the simulations of multi-agent learning with decision markets as discussed in \cite{Wang2022DecisionProblems}, where the principal does not have her own signal. The agents sequentially forecast the principal's signal, conditional on their own signals, and they step-by-step aggregate the available information. After the last agent makes the final forecast, the principal scores all forecasts with a strictly proper scoring rule based on her proxy (i.e., her own signal). Once the principal has learned the correct mapping, she deterministically maps her signal and the final reports to an action. 

We model each agent as a contextual bandit problem with a continuous action space. For each agent, the input constitutes of the signals and the reports from the previous agent. (The first agent uses even odds, or a common prior as input.) The agents will receive a score from the principal that correlates with their contribution to predicting the principal's signal and perform a learning algorithm with it. Unlike in previous work \cite{Wang2022DecisionProblems} the principal is here assumed to learn with policy gradient methods as well, to use her own signal and the forecasts for her own signal to select one of the available actions. This is modelled as contextual bandit problems with discreet action space. Afterwards, the principal observes the Bernoulli outcome of the selected action and updates the policy accordingly.

\subsection{Problem Setup and Notation}
The problem setup of this paper generally follows the work by Wang and Pfeiffer \cite{Wang2022DecisionProblems} with several key differences. The multi-agent contextual bandit problem arises repeatedly and agents can learn from the score from previous rounds. We denote the time step as $T\in\{1,2,\ldots,n\}$. At the beginning of each time step $T$, the principal $P$ faces a finite and discrete action set and selects an action $A\in\{1,2,\ldots,k\}$ from it. The outcome space is a $k$ dimensional Bernoulli vector $\Omega\in\{1,0\}^k$ and the outcome of arm $A$ is denoted as $\Omega_{\left(A\right)}\in\{1,0\}$. The principal desires the actions $A$ with outcome $\Omega_{\left(A\right)}=1$. Each agent $E\in\{1,2,\ldots,m\}$ and the principal receive iid signals about one of the actions $A$. The signals are denoted as $D_{(P,A)}$ for the principal and $D_{(E,A)}$ for agent $E$. The signals are Bernoulli variables, i.e., $D_{\left(P,A\right)}\in\{1,0\}$, and sampled with a stationary probability distribution according to the outcome type of an action. If outcome $\Omega_{(A)}$ for action $A$ is 1, the iid signal $D_{(P,A)}$ or $D_{(E,A)}$ for that action is 1 at probability 2/3 and 0 at probability 1/3. If outcome $\Omega_{(A)}$ for action $A$ is 0, the iid signal $D_{(P,A)}$ or $D_{(E,A)}$ for that action is 1 at probability 1/3 and 0 at probability 2/3. The action space of agents is a multi-dimensional real number $\mathbb{R}^k$, of which $k$ is the cardinality of the principal’s action set. We refer to an agent’s actions as reports, which can be in log-odds or in probabilistic format. We denote the probabilistic reports of agent $E$ as $Pr_{\left(E\right)}\in [0,1]^k$ and the report for a certain principal’s action $A$ as $Pr_{\left(E,A\right)}$, which is the probability that the principal receives a $D_{\left(P,A\right)}=1$ signal about the action $A$. We denote the log-odds report of agent $E$ as $X_{(E,A)}$.

Agents sequentially make reports to forecast the principal’s signal \cite{Hanson2003CombinatorialDesign}, after both the principal and agents receive signals. Agent $E$ receives reports $Pr_{\left(E-1\right)}$ from the previous agent $E-1$ in the sequence. Both $Pr_{\left(E-1\right)}$ and signals  $D_{\left(E\right)}$ composite the contextual vector that we denote as $C_{\left(E\right)}$. The first agent uses even odds instead of received reports. Agent $E$ maintains a policy parameter matrix $\Theta_{\left(E\right)}$ that maps the contextual vector to posterior reports $Pr_{\left(E\right)}$. The posterior reports $Pr_{\left(E\right)}$ are provided to the principal for evaluation, and become part of the subsequent agent's contextual vector. The procedure repeats until the last agent $E=m$ in the sequence provides the final report $Pr_{\left(m\right)}$ to the principal. The ideal final report is essentially an aggregated forecast of the principal’s signal based on all the signals from agents.


In the simulations, we set $k=2$. Agents keep a $6\times2$ policy parameter matrix with random initialisation of weights. The contextual information vector $C_{\left(E\right)}$ is multiplied with the matrix of learning parameters as follows: 
\begin{equation}
    \left(\begin{matrix}c_{r1}\\c_{b1}\\\begin{matrix}c_{p1}\\c_{r2}\\\begin{matrix}c_{b2}\\c_{p2}\\\end{matrix}\\\end{matrix}\\\end{matrix}\right)^\intercal\times\left(\begin{matrix}\theta_{r1}^{\left(1\right)}&\theta_{r1}^{\left(2\right)}\\\begin{matrix}\theta_{b1}^{\left(1\right)}\\\theta_{p1}^{\left(1\right)}\\\begin{matrix}\theta_{r2}^{\left(1\right)}\\\theta_{b2}^{\left(1\right)}\\\theta_{p2}^{\left(1\right)}\\\end{matrix}\\\end{matrix}&\begin{matrix}\theta_{b1}^{\left(2\right)}\\\theta_{p1}^{\left(2\right)}\\\begin{matrix}\theta_{r2}^{\left(2\right)}\\\theta_{b2}^{\left(2\right)}\\\theta_{p2}^{\left(2\right)}\\\end{matrix}\\\end{matrix}\\\end{matrix}\right)=\left(\mu_{(E,1)},\mu_{(E,2)}\right)
\end{equation}

The contextual vector $C_{\left(E\right)}$ has six elements. The element $c_{r1}$ is set to 1 if the agent receives signal 1 for action 1 (i.e., $D_{\left(E,1\right)}=1$); $c_{b1}$ is set to 1 if $D_{\left(E,1\right)}=0$, and similarly $c_{r2}$ and $c_{b2}$ are set to 1 when the received signal is signal $D_{\left(E,2\right)}=1$, and $D_{\left(E,2\right)}=0$, respectively. Above elements remain 0 when there is no signal. The elements $c_{p1}$ and $c_{p2}$ are the prior log-odds transformed probabilistic reports for the first and second actions. The result is a pair of parameters $\left(\mu_{(E,1)},\ \mu_{(E,2)}\right)$, where $\mu_{(E,1)}$ is used to sample a report for action 1, and $\mu_{(E,2)}$ is used to sample a report for action 2. The actual log-odds reports $(X_{(E,1)},X_{(E,2)})$ will be sampled from a normal distribution with the above parameters $\left(\mu_{(E,1)},\ \mu_{(E,2)}\right)$ as means, and a fixed variance to estimate gradients for learning. The probabilistic reports $(Pr_{(E,1)},Pr_{(E,2)})$ will be converted from log-odds reports $(X_{(E,1)},X_{(E,2)})$.

The principal evaluates the probabilistic reports of agent $E$ using a proper scoring rule function and her signal described in \cite{Gneiting2007StrictlyEstimation}:
\begin{equation}
    s:Pr_{\left(E,A\right)}\times D_{\left(P,A\right)}\rightarrow \mathbb{R}
\end{equation}

The reward $R_{\left(E\right)}$ for an agent $E$ is calculated by $s\left(Pr_{\left(E,A\right)},D_{\left(P,A\right)}\right)-s\left(Pr_{\left(E-1,A\right)},D_{\left(P,A\right)}\right)$. As the latter part of the reward is not dependent on the agent $E$; therefore, at time step $T$, agent $E$ optimises the reward by updating the policy parameters with:

\begin{equation}
    \Theta_{\left(E,T+1\right)}=\Theta_{\left(E,T\right)}+\alpha G_{(E,T)}
\end{equation}
where $G_{(E,T)}$ is a matrix of approximated partial derivatives of expected score $\mathbb{E}[s(Pr_{(E,A)},D_{(P,A)})]$ with respect to policy parameters. Because the principal uses a proper score function, the expected score maximises when agent $E$ learns to report $Pr_{\left(E,A\right)}=Pr\left(D_{\left(P,A\right)}\middle| D_{\left(E,A\right)},Pr_{\left(E-1,A\right)}\right)$. The scores for agents depend on agents’ reports and the principal’s signal and, therefore, are decoupled from the principal’s decision-making. The principal can resolve the scores for agents immediately after the aggregated report is available. The scores incentivise the agents to learn to accurately report probabilistic reports for the proxy. 

In the simulations presented here, the principal is a computational model as well, and maintains a policy parameter matrix $\Theta_{\left(P\right)}$. In the simulation, the structure of the policy matrix is the same as the agents’ policy matrix. We denote the contextual vector of the principal as $C_{\left(P\right)}$, which constitutes the principal’s signal $D_{\left(P\right)}$ and the reports from the last agent. While the agent uses the output of the multiplication of the context vector and the weight matrix to sample a report from a normal distribution, the principal uses the output to sample an action using the softmax function. The result $\left(\mu_{(P,1)},\ \mu_{(P,2)}\right)$ can be considered as the preference of an action relative to the other actions. Using the soft-max function, the probabilities of the two actions available in the simulations are given by:

\begin{equation}
    \begin{cases}
        \Phi_1 & =\frac{\exp\left(\mu_{(P,1)}\right)}{\exp\left(\mu_{(P,1)}\right)+\exp\left(\mu_{(P,2)}\right)} \\
        \Phi_2 &=\frac{\exp\left(\mu_{(P,2)}\right)}{\exp\left(\mu_{(P,1)}\right)+\exp\left(\mu_{(P,2)}\right)} \\
    \end{cases}
\end{equation}
The principal sample the action from the distribution $\Phi_{\left(P\right)}=\{\Phi_1,\Phi_2\}$ and executes it. At the time step $T$, after observing the outcome of the executed action, the principal will update its policy parameters to maximise the expectation of the desired outcome to materialise:

\begin{equation}
    \Theta_{\left(P,T+1\right)}=\Theta_{\left(P,T\right)}+\alpha G_{(P,T)}
\end{equation}

$G_{(P,T)}$ is a matrix of approximated partial derivatives of policy parameters with respect to the expected rewards. In other words, the principal learns to interpret the signal $D_{\left(P\right)}$ and the reports from the last agent in sequence $Pr_{\left(m\right)}$ into an action distribution to select the action $A$ that is most likely to lead to an outcome $\Omega_A=1$. Note that the computational model samples the action for approximating gradients, which serves for the learning purpose in our simulations. A principal who knows (or has learned) the relation between final report, her own signal, and outcomes, can use a deterministically select action without changing the incentives for the agents.

\begin{table}[ht] 
\centering
\begin{tabular}{|l|l|l|l|l|l|l|}
\hline
\rowcolor[HTML]{EFEFEF}  
Agent & Action & Signal & Prior       & Posterior   & Decision & Score \\ \hline
1               & 1      & 1      & (0.5,0.5)   & (0.67,0.5)  &          & $-$     \\ \hline
2               & 2      & 1      & (0.67,0.5)  & (0.67,0.67) &          & $0$     \\ \hline
Principal               & 1      & 0      & (0.67,0.67) &             & Action 2        & $+$     \\ \hline
\end{tabular}
\caption{Example decision-making process with two agents. In this example, the outcome for action 1 is 0 and for action 2 is 1. Agent $E=1$ receives a signal $D_{(1,1)}=1$ for action 1 and agent $E=2$ receives a signal $D_{(2,2)}=1$ for action 2. The principal $P$ receives a signal $D_{(P,1)}=0$ for action 1. Assuming the two agents and the principal are well-trained. Agent $E=1$ updates the prior $(0.5,0.5)$ to generate a posterior report $(0.67,0.5)$. Agent $E=2$ does not change the report for action 1 but increases the report for action 2 and generates the posterior report $(0.67,0.67)$. The principal decides to execute action $A=2$ because her signal breaks the tie in favour of action $A=2$. This is a correct decision and that is rewarded by $\Omega_{(2)=1}$. While the expected payoff of agent 1 is positive in this case, agent 1 receives a negative score because it makes the reports for the principal's signal less accurate. Agent 2 receives a score of 0 because the principal receives a signal for action $A=1$ rather than action $A=2$. }
\label{tab:simple_two_cases}
\end{table}

In the simulation, we use a Brier scoring rule. Any proper scoring rules will have consistent results here \cite{Gneiting2007StrictlyEstimation,Hanson2007LogarithmicAggregation}. Both the principal and the agents use an experience replay buffer technique to speed up the training process \cite{Lin1993ReinforcementNetworks,Mnih2013PlayingLearning}. For agent $E$, a tuple $\left(C_{\left(E,T\right)},\mu_{\left(E,T\right)},X_{\left(E,T\right)},R_{\left(E,T\right)}\right)$ consists of the contextual vector, the mean log-odds, the actual reports and the reward at time step $T$ is called a piece of experience. For the principal $P$, the experience tuple $\left(C_{\left(P,T\right)},\Phi_{\left(P,T\right)},\Omega_{\left(A,T\right)},A_T\right)$ consists of the contextual vector, the action distribution and the materialised Bernoulli outcome (which is essentially a 1-0 reward). In each time step, the principal and agents randomly draw a mini-batch of pieces of experience from the experience reply buffer. We assume the experience tuple at time step $I$ is within the mini-batch sampled at time step $T$. The gradient for the principal can be obtained by:
\begin{equation} \label{eq:principal_gradient}
    G_{\left(P,I\right)}=
    \begin{cases}
    & C_{\left(P,I\right)}\left(\Omega_{\left(A,I\right)}-B_{\left(P,I\right)}\right)\left(1-\Phi_{\left(P,I\right)}\left(A_I\right)\right) \\
    & -C_{\left(P,I\right)}\left(\Omega_{\left(A,I\right)}-B_{\left(P,I\right)}\right)\left(\Phi_{\left(P,I\right)}\left(a\right)\right),\forall a\neq A_I \\
    \end{cases}
\end{equation}

The gradient for the agents can be obtained by:
\begin{equation} \label{eq:expert_gradient}
    G_{\left(E,I\right)}=C_{\left(E,I\right)}\left(R_{\left(E,I\right)}-B_{\left(E,I\right)}\right)\frac{X_{\left(E,I\right)}-\mu_{\left(E,I\right)}}{\sigma^2}
\end{equation}
In our simulation, $C_{(P,I)}$ and $C_{(E,I)}$ have $6\times1$ dimension and $\Phi_{(P,I)}$, $X_{(E,I)}$ and $\mu_{(E,I)}$ have $1\times2$ dimension. Therefore, the Gradients have the same dimensionality as the policy matrix. We use a different time step notation $I$ to emphasise that at time step $T$, the training does not necessarily use the experience from time step $T$. In both equations \ref{eq:principal_gradient} and \ref{eq:expert_gradient}, $B_{\left(P,I\right)}$ and $B_{\left(E,I\right)}$ are two baselines functions that do not vary with the action. The baseline function should have no effect on the expectation but influences the variance and will speed up the training process. A popular choice of the baseline function is a running average of the reward. Finally, both the principal and agents will learn with the average of the mini-batch gradients.

In this work, we set a Bayesian updating model as a benchmark. At the first stage, the Bayesian model receives all the same signals $D_{\left(E\right)}$ as the agents. The Bayesian model generates a posterior according to the agents’ signals $\widehat{Pr_{\left(A\right)}}=Pr\left(D_{\left(P,A\right)}\middle|D_{\left(1,A\right)},\ldots,D_{\left(m,A\right)}\right)$. As once the accessibility of distributed and proprietary signals is allowed, the problem become trivial. Therefore, we consider the Bayesian update posterior as the ideal report that the agents can generate. We evaluate the mean squared error between the aggregated report and the Bayesian updated posterior:
\begin{equation} \label{eq:peer_decision_error}
    Er=\sum_{A}\left(Pr_{\left(m,A\right)}-\widehat{Pr_{\left(A\right)}}\right)^2
\end{equation}

To evaluate performance in decision making, we use a Bayesian model that will take both the principal’s signal and the agents’ signals into consideration and compute the posterior to $\widehat{Pr_{\left(A\right)}^\prime}=Pr\left(\Omega_{\left(A\right)}\middle|D_{\left(1,A\right)},\ldots,D_{\left(m,A\right)},D_{\left(P,A\right)}\right)$. We deterministically select the action A that has the highest chance to have outcome $\Omega_{\left(A\right)}=1$ according to $\widehat{P{r^\prime}_{\left(A\right)}}$, and record the frequency at which this procedure selects an action with outcome 1. For comparison, we also record the average frequency that the principal selected action’s outcome to be 1.

\subsection{Simulation Results} \label{sec:pd_simulation_results}
This section will show the result of a multi-agent bandit learning system that employs our mechanism for deterministic decision-making problems in the setting we mentioned in the previous section.

\begin{figure}[ht]
    \centering
    \includegraphics[width=\textwidth]{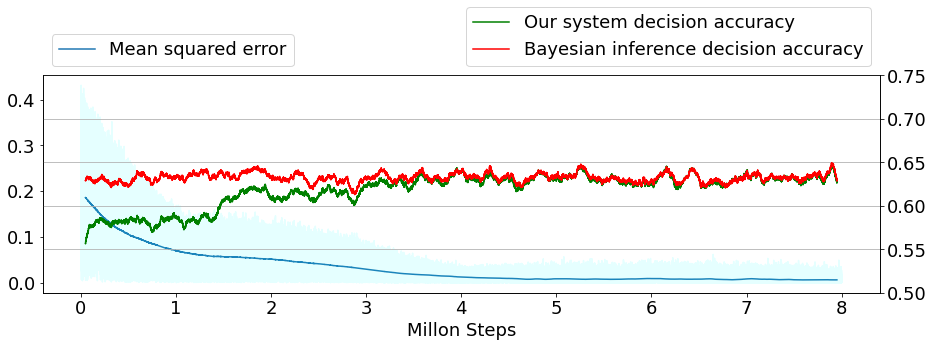}
    \caption{System performance of a three-agent system and the Bayesian inference. The blue line is the mean squared error $Er$ between the aggregated report $Pr_{(m,A)}$ from the market and a Bayesian inference $\widehat{Pr_{\left(A\right)}}$ that uses the same information. The light cyan marks the range of the actual error. The red and green lines are the average times the selected action’s outcome that the principal desires.}
    \label{fig:peer_decision_loss}
\end{figure}

Figure \ref{fig:peer_decision_loss} shows that the multi-agent system's error, calculated by equation \ref{eq:peer_decision_error}, converges after around 5 million steps. After the agents’ policy parameters converge, the decision quality of the principal is as good as a Bayesian inference model that can access all the signals. The result suggests that our mechanism achieves the most informed decision-making with a deterministic decision rule.

\section{Conclusion and Discussion}
This research proposes a set of closely related mechanisms to address a collective decision-making problem. That is, a principal wants to decide between several actions, but most information about the quality of the actions is distributed and privately owned by self-interested agents. To make an informed decision, the principal needs to elicit the information with incentives and aggregate it in a suitable way for subsequent decision-making. Existing collective decision-making mechanisms face a dilemma: either decision cannot be made deterministically without triggering manipulative behaviours, or the information cannot be elicited from more than one agent to make a deterministic recommendation \cite{Othman2010DecisionMarkets,Chen2014ElicitingMaking}. 

Our mechanisms address information elicitation and aggregation under a deterministic decision rule through the prediction of a principal's proxy. This approach resembles peer prediction mechanisms, in that rather than predicting future outcomes, our mechanism requires agents to predict the nature of the principal's proxy, which is statistically correlated with the future outcome given an action. The principal then can use a deterministic decision rule to map the aggregated report for the proxy to an action.

We further expand this mechanism to two additional variants. The first expansion lifts the assumption that the principal requires a proxy herself. The mechanism separates an agent from the rest as an advisor. The principal can elicit the proxy from the advisor by providing a share of her reward to align their interest. The second expansion employs a peer prediction mechanism and separates two agents to predict the signal that the other party possesses. As long as one chooses a proper peer prediction mechanism, the elicited signals are as if the principal owned the information. In both variants, the information of the other agents is aggregated and used to make a deterministic decision. Note the principal must ensure that collusion does not occur between the advisor and the other agents, or the two peers and the other agents. 

We use a multi-agent contextual bandit system to simulate the dynamics of learning for our first mechanism. The result shows that the average frequency of the selected action's outcome desired by the principal using our mechanism is as high as under a Bayesian inference model with identical information. It suggests our mechanism solves the collective decision-making problem with a deterministic decision rule.

There are several future study directions. An important direction is to employ the mechanisms described here to solve a contextual bandit problem that involves multiple agents who are unwilling to share raw signals, i.e., user profiles. Such a scenario fits into a growing research direction of federated learning. Federated learning, unlike the traditional supervised learning model, forbids a centralised party from directly collecting data for learning \cite{Konecny2016FederatedEfficiency,Yang2019FederatedApplications}. Federated bandit problems fall into this category, which train recommendation systems without accessing private user data. Our mechanisms inherently do not require accessing the agents’ signals and potentially provide another direction to solve the federated bandit problems.

In the mechanisms presented here, we distinguish between the principal and agents. For many peer decision problems, there is no such distinction. All agents could, in principle, receive signals and get into the position to execute an action. We can expand our mechanism to such a peer decision problem by randomly selecting an agent as the principal at each time step, and the rest of the procedure follows the first mechanism. This mechanism is most useful when the peers have no conflicting preferences for the actions’ outcome. For instance, all agents prefer success over failure as the selected action’s outcome. When the agents are not picked as the principal, they are self-interested in that they seek to profit from their signals.

We validate our mechanism with self-interested computational agents. It remains to be studied how well the mechanism works for a more complex action and outcome space. It is also worth investigating how well the mechanism works, for instance, with human subjects in a laboratory setting, and how it compares to alternative mechanisms of collective decision making. 

\bibliography{decision_markets,multiagent_system,bandit_rl,federated_learning,mrl,peer_prediction}             
\bibliographystyle{plainnat} 
\end{document}